\documentclass[12pt]{article}
\usepackage{esfconf}
\usepackage{amsmath}
\usepackage{amsfonts}
\usepackage{amssymb}
\usepackage{cite}
\def\1ad{\mbox{\normalsize $^a$}}
\def\2ad{\mbox{\normalsize $^b$}}
\def\3ad{\mbox{\normalsize  $^c$}}
\newcommand{\teight}[1]{t_8^{(#1)}}

\begin{document}
{\flushright{\small DAMTP-2000-91,\quad NORDITA-2000/87 HE,\quad SPHT-T00/117,\quad \tt hep-th/0010182\\}}
\vskip .6cm

\title{
Supersymmetric $R^4$ actions and quantum corrections to superspace
torsion constraints
}

\authors{K. Peeters\adref{a}, P. Vanhove\adref{b} and A. Westerberg\adref{c}}

\addresses{\1ad CERN, TH-division 1211 Geneva 23, Switzerland
\nextaddress
\2ad SPHT, Orme des Merisiers, CEA / Saclay, 91191 Gif-sur-Yvette, France
\nextaddress
\3ad NORDITA, Blegdamsvej 17, DK-2100 Copenhagen \O, Denmark
\nextaddress
\tt k.peeters, p.vanhove@damtp.cam.ac.uk, a.westerberg@nordita.dk}


\maketitle


\begin{abstract}
  We present the supersymmetrisation of the anomaly-related $R^4$ term
  in eleven dimensions and show that it induces no non-trivial
  modifications to the on-shell supertranslation algebra and the
  superspace torsion constraints before inclusion of gauge-field
  terms.\footnote{Based on talks given by K.P. at the SPG meeting in
    Cambridge (February 2000) and in Groningen (September 2000),
    by A.W. at the Nordic Network Meeting in Copenhagen (May 2000),
    and by P.V. at the Fradkin Memorial
    Conference in Moscow (June 2000) and at the ARW Conference in Kiev
    (September 2000).}
\end{abstract}


\section{Higher-derivative corrections and supersymmetry}

The low-energy supergravity limits of superstring theory as well as
D-brane effective actions receive infinite sets of correction terms,
proportional to increasing powers of $\alpha'=l_s^2$ and induced by
superstring theory massless and massive modes.  At present,
eleven-dimensional supergravity lacks a corresponding microscopic
underpinning that could similarly justify the presence of
higher-derivative corrections to the classical Cremmer-Julia-Scherk
action \cite{CremmerJS}. Nevertheless, some corrections of this kind
are calculable from unitarity arguments and super-Ward identities in
the massless sector of the theory \cite{Bern} or by anomaly
cancellation arguments \cite{DuffLM, VafaW}.

Supersymmetry puts severe constraints on higher-derivative corrections.
For example, it forbids the appearance of certain corrections (like, e.g, 
$R^3$ corrections to supergravity effective actions \cite{Grisaru}), 
and groups terms into various invariants 
\cite{BergshoeffR,deRooSW,Suelmann,GreenS}. The structure of the invariants
that contain anomaly-cancelling terms is of great importance due to 
the quantum nature of the anomaly-cancellation mechanism and is the
main concern of this note.

Higher-derivative additions to the supergravity actions are in general
compatible with supersymmetry only if the transformation rules for the
fields also receive higher-derivative corrections:
\begin{equation}
\bigg( \delta_0 + \sum_n (\alpha')^n \delta_n \bigg) \bigg( S_0 + \sum_n
(\alpha')^n S_n \bigg) = 0\, .
\end{equation}
As a consequence, the field-dependent structure coefficients on the 
right-hand side of the supersymmetry algebra,
\begin{equation}
\label{e:algebra}
{}[\delta^{\text{susy}}_1,\delta^{\text{susy}}_2]=
\delta^{\text{translation}} +  \delta^{\text{susy}} +
\delta^{\text{gauge}} +  \delta^{\text{Lorentz}} \, , 
\end{equation}
will be modified as well. When the theory is formulated in superspace
the structure of the algebra is related to the structure of the
tangent bundle, the link being provided by the constraints on the
superspace torsion. In particular, corrections to the parameters
modify the superspace constraints. However, since some corrections are
reabsorbable by suitable rotations of the tangent bundle basis, 
not all corrections are physical.

We report here on the supersymmetrisation of the anomaly-related terms
$(\alpha')^2 B\wedge F^4$ for super-Maxwell theory coupled to $N$=1
supergravity in ten dimensions and $(\alpha'_M)^3 C\wedge t_8 R^4$
(where $(\alpha'_M)^3 =4\pi\, (l_P)^6$) in eleven dimensions performed
in \cite{PeetersVW}. While the former does not require any corrections
to the superspace constraints, there are strong indications based on a
previous superspace analysis~\cite{Howe} that the latter does induce
such modifications. However, we show that there are no such
corrections which are proportional to (powers of) the Weyl tensor only.  We
present here only the more salients aspect of the analysis and refer
to the article \cite{PeetersVW} for computational and bibliographical
details.

Our main motivation to look for non-trivial corrections to superspace 
constraints comes from the link between these constraints and the
kappa symmetry of M-branes \cite{BergshoeffST,BandosLNPST,AganagicPPS} 
and D-branes \cite{CederwallGNSW,BergshoeffT,AganagicPS}.
Classical kappa invariance of the M- and D-brane world-volume actions 
--- a key requirement for these objects to be supersymmetric --- imposes 
the on-shell constraints on the background superspace supergravity fields, 
among them the superspace torsion. For this reason, any non-trivial 
modification to the constraints is expected to require new terms in
the world-volume actions for the branes in order for kappa symmetry to
be preserved.

\section{Construction of an abelian $F^4$ superinvariant in D=10}

As a first step in our analysis of the implications of higher-derivative
corrections to the supersymmetry algebra, we discuss the construction
of the abelian $(\alpha')^2 (t_8 F^4 - B\wedge F^4)$ for $N$=1
super-Maxwell theory coupled to gravity in ten dimensions.

The field content of the on-shell super-Maxwell theory comprises an
abelian vector $A_\mu$ and a negative-chirality Majorana-Weyl spinor
$\chi$. Since we are interested in local supersymmetry invariance we
have to take into account also the interactions with the zehnbein 
$e_\mu{}^r$, the negative-chirality Majorana-Weyl gravitino $\psi_\mu$ 
and the two-form $B_{\mu\nu}$ from the supergravity multiplet.
The classical action (leaving out the gravitational sector)
\begin{equation}
\label{e:superF2}
S_{F^2} = \int\!{\rm d}^{10}x \,e \big[ -\tfrac{1}{4}
F_{\mu\nu}F^{\mu\nu} - 8\,\bar\chi \not\!\!D (\omega) \chi
+2\,\bar\chi\Gamma^\mu \Gamma^{\nu\rho} \psi_\mu\, F_{\nu\rho}\, \big]
\end{equation}
is invariant under the local supersymmetry transformations
\begin{equation}
\label{e:YMtrafo}
\delta A_\mu = -4\,\bar\epsilon \Gamma_\mu \chi\, , \qquad
 \delta  \chi = \tfrac{1}{8}\Gamma^{\mu\nu}\epsilon\, F_{\mu\nu}\, .
\end{equation}
For local supersymmetry we have to consider the transformations of the
supergravity multiplet fields as well (neglecting terms proportional to 
the two-form $B_{\mu\nu}$ and the corresponding field strength, 
$H_{\mu\nu\rho}$):
\begin{equation}
\label{e:SYMtrafos}
\delta e_\mu{}^r = 2\,\bar\epsilon\Gamma^r\psi_\mu\,, \quad 
\delta \psi_\mu = D_\mu(\omega) \epsilon + \cdots\,,\quad 
\delta B_{\mu\nu} = 2\, \bar\epsilon\Gamma_{[\mu}
\psi_{\nu]}\, .
\end{equation}
The $F^4$ action invariant under the local supersymmetry
transformations listed above has been determined
previously~\cite{deRooSW,Suelmann} and extends the globally
supersymmetric action of~\cite{MetsaevR,BergshoeffRS}. Using
additional string input, we have managed to group all terms (including
the fermionic bilinears) in a very compact way using the well-known
$t_8$-tensor~\cite{PeetersVW}.  The result is
\begin{equation}
\label{e:SYMstring}
\begin{aligned}
S_{F^4}& = \frac{(\alpha')^2}{32}\int\! {\rm d}^{10}x \Big[ \,\tfrac{1}{6}
e\, \teight{r} F_{r_1r_2}\cdots F_{r_7r_8}
 + \tfrac{1}{12} \varepsilon_{10}^{(r)} B_{r_1r_2}
F_{r_3r_4} \cdots F_{r_9r_{10}} \\[1ex]
& - \tfrac{32}{5} e\, \teight{r} \eta_{r_2r_3} (\bar\chi \Gamma_{r_1}
D_{r_4}(\omega)\chi)  F_{r_5r_6}F_{r_7r_8}
 + \tfrac{12\cdot32}{5} e\, (\bar\chi\Gamma_{r_1}D_{r_2}(\omega)\chi)
F^{r_1m}F_m{}^{r_2}  \\[1ex]
&  -\tfrac{16}{5!} \varepsilon_{10}^{(r)} (\bar\chi \Gamma_{r_1\cdots r_5}
 D_{r_6}(\omega)\chi) F_{r_7r_8} F_{r_9r_{10}} 
  + \tfrac{16}{3}e\,\teight{r} (\bar\psi_{r_1} \Gamma_{r_2}\chi)
F_{r_3r_4}F_{r_5r_6}F_{r_7r_8} \\[1ex]
& + \tfrac{8}{3}e\, (\bar\psi_{m}\Gamma^{mr_1\cdots r_6} \chi) F_{r_1
r_2}\cdots F_{r_5r_6} \Big] \,.
\end{aligned}
\end{equation}
The local supersymmetry invariance of the combined action $S_{F^2}+ S_{F^4}$ 
requires that the supersymmetry transformations be modified according to
($F^2:=F^{mn}F_{nm}$)
\begin{equation}
\label{e:modtrafoYM}
\begin{aligned}
\delta A_\mu &= -4\,\bar\epsilon \Gamma_\mu \chi - (\alpha')^2\Big[ 
\tfrac{1}{4}(\bar\epsilon\Gamma_\mu\chi) F^2 - (\bar\epsilon\Gamma^m\chi) 
F^2_{m\mu} - \tfrac{1}{8}(\bar\epsilon\Gamma^{r_1\cdots r_4}{}_\mu\chi) 
F_{r_1r_2}F_{r_3r_4}\Big]\, ,\\[1ex]
\delta \chi &= \tfrac{1}{8}\Gamma^{\mu\nu}\epsilon\, F_{\mu\nu} 
+ \tfrac{1}{768} (\alpha')^2\Big[ \teight{r} \Gamma_{r_7r_8}\epsilon 
- \Gamma^{r_1\cdots r_6}\epsilon \Big] F_{r_1r_2}F_{r_3r_4} F_{r_5r_6}\,.
\end{aligned}
\end{equation}
It can be verified that the structure of the supersymmetry algebra is
not modified by the order-$(\alpha')^2$ corrections \cite{MetsaevR,BergshoeffRS,PeetersVW}:
\begin{equation}
{}\big[ \delta_{\epsilon_1}^{(\alpha')^0} + \delta_{\epsilon_1}^{(\alpha')^2},
\delta_{\epsilon_2}^{(\alpha')^0} + \delta_{\epsilon_2}^{(\alpha')^2}
\big] A_\mu = {}\big[ \delta_{\epsilon_1}^{(\alpha')^0},
\delta_{\epsilon_2}^{(\alpha')^0} \big] A_\mu + {\cal
  O}\big((\alpha')^4\big)\, . 
\end{equation}
Consequently, the structure of the superspace torsion constraints will be 
the same as for the classical theory to this order. This observation is 
related to the fact that it is possible to supersymmetrise the 
Dirac-Born-Infeld actions while imposing only the classical constraints 
\cite{BaggerG}.  

\section{Construction of the $C\wedge R^4$ superinvariant in D=11}

Noticing the close parallel between the classical supersymmetry
transformations for the super-Maxwell and the supergravity fields
\begin{equation}
\begin{aligned}
\label{e:trafo_comp}
\delta\chi &= \tfrac{1}{8}\Gamma^{\mu\nu}\epsilon \, F_{\mu\nu}\,, 
&\qquad\qquad   
\delta\psi_{rs} &= \tfrac{1}{8} \Gamma^{\mu\nu}\epsilon \, R_{\mu\nu rs}
+ \cdots \,,\\[1ex]
\delta F_{\mu\nu} &= - 8 D_{[\mu} (\bar\epsilon\Gamma_{\nu]}\chi) \,,  &
\delta R_{\mu\nu}{}^{rs} &= -8 D_{[\mu}(\bar\epsilon\Gamma_{\nu]}\psi^{rs}) 
 \,\\[1ex]
&&&+ 4 D_{[\mu}(\bar\epsilon\Gamma_{\nu]}\psi^{rs}
+2\,\bar\epsilon\Gamma^{[r}\psi^{s]}{}_{\nu]}) + \cdots \,,
\end{aligned}
\end{equation}
it is tempting to make the following substitution in the super-Maxwell 
action:
\begin{equation}
\label{e:SYM2SUGRA}
F_{r_1r_2}  \rightarrow R_{r_1r_2s_1s_2},\quad
\chi        \rightarrow \psi_{s_1s_2} ,\quad
D_r\chi \rightarrow D_r\psi_{s_1s_2}\, ,
\end{equation}
Unfortunately, the difference in structure between the equations of
motion for the gauge potential and the spin connection implies that
the previous mapping does not commute with supersymmetry, as can be
seen by the presence of the second line in the supersymmetry
transformation of the Riemann tensor above.
Another crucial difference between the super-Maxwell and supergravity 
cases is that, when subtracting all the lowest-order equations of motions, 
it is necessary to make the following substitution for the Riemann tensor:
\begin{equation}
\label{e:Riemann_subst}
R_{mn}{}^{pq} \rightarrow W_{mn}{}^{pq} - \frac{16}{d-2}
\delta_{[m}{}^{[p}(\bar\psi_{|r|}\Gamma^{|r|}\psi_{n]}{}^{q]}
- \bar\psi^{|r|}\Gamma^{q]}\psi_{n]r}) \, .
\end{equation}
Taking all these facts into account, as well as the information from
string-amplitude analysis that the extra $s$-type indices
in~(\ref{e:SYM2SUGRA}) should be contracted with an additional
$t_8^{(s)}$ tensor, we arrive at the following M-theory $C\wedge R^4$
invariant after lifting to eleven dimensions \cite{PeetersVW}:
\begin{equation}
\label{e:elevendimI_X}
\begin{aligned}
%
%
(\alpha_M')^{-3} {\cal L}_{\Gamma^{[0]}} =\,\, &+&\tfrac{1}{192}e\,
\teight{r}&\teight{s}\, W_{r_1r_2s_1s_2}\cdots W_{r_7r_8s_7s_8} \\[1ex]
&+&\tfrac{1}{(48)^2}&\varepsilon^{t_1t_2t_3r_1\cdots r_8}\,
\teight{s} C_{t_1t_2t_3}
W_{r_1r_2s_1s_2}\cdots W_{r_7r_8s_7s_8}\, , \\[1ex]
%
%
(\alpha_M')^{-3} {\cal L}_{\Gamma^{[1]}} =\,\, &-&4\,e\,\teight{s} 
&(\bar\psi_{s_1s_2} \Gamma_{r_1} D_{r_2} \psi_{s_3s_4})
W_{r_1r_3s_5s_6}W_{r_3r_2s_7s_8} &\quad&  \\[1ex]
&-& \tfrac{1}{4}e\,\teight{s} &(\bar\psi_{r_1}\Gamma_{r_2}\psi_{s_7s_8}) 
W_{r_1r_2s_1s_2} W_{mns_3s_4}W_{nms_5s_6} & & \\[1ex]
&-& e\,\teight{s}& (\bar\psi_{r_1}\Gamma_{r_2}\psi_{s_7s_8}) 
W_{r_1ms_1s_2}W_{mns_3s_4}W_{nr_2s_5s_6} \\[1ex]
 &+& e\,\teight{s} &(\bar\psi_{r_1}\Gamma_{s_7}\psi_{r_2s_8}) W_{r_1r_2s_1s_2}
W_{mns_3s_4}W_{nms_5s_6} \\[1ex]
 &-& 4\,e\,\teight{s}&(\bar\psi_{r_1}\Gamma_{s_7}\psi_{r_2s_8}) 
W_{r_1ms_1s_2}W_{mns_3s_4}W_{nr_2s_5s_6} \\[1ex]
&+&\tfrac{2}{9} e\,\teight{s} &(\bar\psi_m\Gamma_n\psi_{ms_8}) 
W_{pqs_1s_2}W_{qps_3s_4}W_{ns_7s_5s_6} \\[1ex]
&-& \tfrac{8}{9} e\,\teight{s} &(\bar\psi_m\Gamma_n\psi_{ms_8}) 
W_{nps_1s_2}W_{pqs_3s_4}W_{qs_7s_5s_6} \, , \\[1ex]
%
%
(\alpha_M')^{-3} {\cal L}_{\Gamma^{[3]}} =\,\, &+&2\,e\, \teight{s} &(\bar\psi_{s_
5s_6}\Gamma_{r_1r_2r_3} D_{r_4}
\psi_{s_7s_8}) W_{r_1r_2s_1s_2}W_{r_3r_4s_3s_4}  \\[1ex]
 &-& \tfrac{1}{8}e\,\teight{s} &(\bar\psi_m \Gamma^{mr_1r_2}\psi_{s_7s_8})
W_{r_1r_2s_1s_2} W_{pns_3s_4}W_{nps_5s_6} & &  \\[1ex] 
 &+& \tfrac{1}{2}e\, \teight{s} &(\bar\psi_m \Gamma^{mr_1r_2} \psi_{s_7s_8}) 
W_{r_1ps_1s_2}W_{pns_3s_4}W_{nr_2s_5s_6} \\[1ex]
 &+&  e\,\teight{s} &(\bar\psi_m \Gamma^{r_1r_2r_3}\psi_{s_7s_8}) 
W_{r_1r_2s_1s_2} W_{mn s_3s_4}W_{nr_3 s_5s_6}\, , \\[1ex]
%
%
(\alpha_M')^{-3} {\cal L}_{\Gamma^{[5]}} =\,\,  &+&\tfrac{1}{8}e \,
  \teight{s} &(\bar\psi^{r_6} \Gamma^{r_1\cdots r_5}\psi_{s_7s_8})
  W_{r_1r_2s_1s_2}W_{r_3r_4s_3s_4} W_{r_5r_6s_5s_6}  \, , \\[1ex]
%
%
(\alpha_M')^{-3} {\cal L}_{\Gamma^{[7]}} =\,\,  &+&\tfrac{1}{48}e \,
  \teight{s}&(\bar\psi_m \Gamma_{m r_1\cdots r_6}\psi_{s_7s_8})
  W_{r_1r_2s_1s_2}W_{r_3r_4s_3s_4} W_{r_5r_6s_5s_6}  \, . 
\end{aligned}
\end{equation}
Even though the elfbein supersymmetry transformation rule receives
$(\alpha'_M)^3$ modifications, by computing the closure of the
supersymmetry algebra~(\ref{e:algebra}), we find \cite{PeetersVW} that
the translation parameter does \emph{not} receive corrections that
cannot be absorbed by field redefinitions.

\section{Superspace approach}

It can be argued that in the completely general ansatz for the
dimension-zero torsion constraint
\begin{equation}
\label{e:quantumtorsion}
T_{ab}{}^r = 2\big( \left({\cal C}\Gamma^{r_1}\right)_{ab}
X^r{}_{r_1} +\tfrac{1}{2!} \left({\cal
    C}\Gamma^{r_1r_2}\right)_{ab} \, X^r{}_{r_1r_2}
+\tfrac{1}{5!} \left({\cal C}\Gamma^{r_1\cdots
    r_5}\right)_{ab} \,X^r{}_{r_1\cdots r_5}\big)\, , 
\end{equation}
the coefficient $X^r{}_{r_1}$ can be set equal to $\delta^r{}_{r_1}$
and all fully antisymmetric tensors contained in $X^r{}_{r_1r_2}$ and
$X^r{}_{r_1\cdots r_5}$ can be set to zero by a choice of tangent
bundle basis (see, e.g., \cite{CederwallGNN}). This leaves as the only
candidates for non-trivial M-theory corrections to the standard
dimension-zero constraint
\begin{equation}
\label{e:classical}
T_{ab}{}^r =  2\left({\cal C}\Gamma^r\right)_{ab} 
\end{equation}
the SO(1,10) representations {\bf 429} and {\bf 4290} of the $\Gamma_{[2]}$ 
and $\Gamma_{[5]}$ coefficients, respectively.  Therefore, from the 
component-field analysis of the previous section we conclude that the 
higher-order invariant~(\ref{e:elevendimI_X}) does not induce any 
modifications to the torsion constraint~(\ref{e:quantumtorsion}).

Howe has shown in~\cite{Howe} that by imposing \emph{only} the
constraint~(\ref{e:classical}), the full classical, on-shell,
eleven-dimensional supergravity theory of \cite{CremmerJS} follows.
Hence, we conclude from the absence of corrections
to~(\ref{e:quantumtorsion}) induced by the $R^4$
invariant~(\ref{e:elevendimI_X}), that any non-trivial M-theory
corrections to the classical supergravity theory requires the
inclusion of the four-form field strength. In addition, one may get
such corrections from the inclusion of the $\epsilon\epsilon R^4$
interaction, which at the level of our analysis is part of a separate
superinvariant and has therefore not been taken into account yet.

One can argue, on the basis of lifting of the type IIA
action~\cite{KP}, that this $\epsilon\epsilon R^4$ term should be
present in the eleven-dimensional theory as well, but it is
interesting to note that our results give another strong indication
that this term should be present. Otherwise our analysis, when
combined with Howe's, would imply that the dynamics encoded in the
action~(\ref{e:elevendimI_X}) does not correspond to any non-trivial
corrections to the classical supergravity theory of~\cite{CremmerJS}
for configurations with vanishing four-form field strength.

In this context, let us also mention that in parallel with our 
component-field based approach to uncover the superspace underlying 
M-theory, a complementary line of attack based on an analysis of the 
superspace Bianchi identities has been initiated by Cederwall et 
al.~in~\cite{CederwallGNN}.


\noindent{\bf Acknowledgements.} K.P. and P.V. are supported by PPARC
grant\break PPA/G/S/1998/00613. P.V. thanks the TMR contract ERB FMRXCT
96-0012, for financial support.


{\raggedright
}

\end{document}